\documentclass[preprint,preprintnumbers,amsmath,amssymb,dvipdfmx]{revtex4}
\usepackage[dvipdfmx]{graphicx}
\usepackage{mathptmx,bm}
\usepackage[compatibility=false]{caption}
\usepackage{color}
\usepackage{lineno}

\begin{document}


\title{Band-dependent superconducting gap in SrFe$_{2}$(As$_{0.65}$P$_{0.35}$)$_{2}$ studied by angle-resolved photoemission spectroscopy} 

\author{H. Suzuki$^{1}$, T. Kobayashi$^{2}$, S. Miyasaka$^{2,3}$, K. Okazaki$^{1,4}$, T. Yoshida$^{1,3}$, M. Horio$^{1}$, L. C. C. Ambolode II$^{1}$, Y. Ota$^{4}$, H. Yamamoto$^{4}$, S. Shin$^{3,4}$, M. Hashimoto$^5$, D. H. Lu$^5$, Z.-X. Shen$^5$, S. Tajima$^{2,3}$ and A. Fujimori$^{1,3}$}
 
\affiliation{$^1$Department of Physics, University of Tokyo,
Bunkyo-ku, Tokyo 113-0033, Japan}

\affiliation{$^2$Department of Physics, Osaka University, Toyonaka, Osaka 560-8531, Japan}

\affiliation{$^3$JST, Transformative Research-Project on Iron
Pnictides (TRIP), Chiyoda, Tokyo 102-0075, Japan}

\affiliation{$^{4}$Institute for Solid State Physics (ISSP), University of Tokyo, Kashiwa, Chiba 277-8581, Japan}

\affiliation{$^5$Stanford Synchrotron Radiation Lightsource, SLAC National Accelerator Laboratory, Menlo Park, California 94305, USA}

\date{\today}

\begin{abstract}

The isovalent-substituted iron pnictide compound SrFe$_{2}$(As$_{1-x}$P$_{x}$)$_{2}$ exhibits multiple evidence for nodal superconductivity via various experimental probes, such as the penetration depth, nuclear magnetic resonance and specific heat measurements. The direct identification of the nodal superconducting (SC) gap structure is challenging, partly because the presence of nodes is not protected by symmetry but instead caused by an accidental sign change of the order parameter, and also because of the three-dimensionality of the electronic structure. We have studied the SC gaps of SrFe$_{2}$(As$_{0.65}$P$_{0.35}$)$_{2}$ in three-dimensional momentum space by synchrotron and laser-based angle-resolved photoemission spectroscopy. The three hole Fermi surfaces (FSs) at the zone center have SC gaps with different magnitudes, whereas the SC gaps of the electron FSs at the zone corner are almost isotropic and $k_{z}$-independent. We propose that the SC gap of the outer hole FS changes sign around the Z-X [($0, 0, 2\pi$)-($\pi, \pi, 2\pi$)] direction. 
\end{abstract} 
\maketitle

\newcounter{Fig1}
\setcounter{Fig1}{1}
\newcounter{Fig2}
\setcounter{Fig2}{2}
\newcounter{Fig3}
\setcounter{Fig3}{3}
\newcounter{Fig4}
\setcounter{Fig4}{4}
\newcounter{Fig5}
\setcounter{Fig5}{5}
\newcounter{Fig6}
\setcounter{Fig6}{6}
\newcounter{Fig7}
\setcounter{Fig7}{7}

\section*{Introduction}
Since the discovery of high-temperature superconductivity in the iron pnictides and chalcogenides \cite{Kamihara.Y_etal.Journal-of-the-American-Chemical-Society2008}, the mechanism of Cooper pairing and the symmetry of the order parameter have been central issues of debate \cite{Hirschfeld.P_etal.Reports-on-Progress-in-Physics2011,Chubukov.AAnnual-Review-of-Condensed-Matter-Physics2012}. In the spin-fluctuation-mediated superconductivity, intra-orbital nesting between hole and electron Fermi surfaces (FSs) brings about a sign-changing $s^{\pm}$-wave superconducting (SC) state \cite{Kuroki.K_etal.Phys.-Rev.-Lett.2008,Mazin.I_etal.Phys.-Rev.-Lett.2008,Scalapino.D_etal.Rev.-Mod.-Phys.2012}. In the orbital-fluctuation-mediated superconductivity, a sign-preserving $s^{++}$-wave SC state is expected \cite{Kontani.H_etal.Phys.-Rev.-Lett.2010,Onari.S_etal.Phys.-Rev.-Lett.2012}. In heavily overdoped systems such as Ba$_{1-x}$K$_{x}$Fe$_{2}$As$_{2}$, where the electron FSs at the zone corner are absent or much smaller than the hole FSs at the zone center, a $d$-wave pairing \cite{Thomale.R_etal.Phys.-Rev.-Lett.2011} and time-reversal-symmetry-broken pairings such as $s+is$ and $s+id$ symmetries have been proposed \cite{Maiti.S_etal.Phys.-Rev.-B2013,Lin.S_etal.Phys.-Rev.-B2016}. The diversity of the SC gap structure among various materials and doping elements characterizes the FeSCs in comparison with the cuprates \cite{Scalapino.D_etal.Phys.-Rep.1995}.

Among the FeSCs, the isovalently-substituted BaFe$_{2}$(As$_{1-x}$P$_{x}$)$_{2}$ (Ba122P) system \cite{Kasahara.S_etal.Phys.-Rev.-B2010} has attracted particular attention since the presence of line nodes in the SC order parameter was proposed from NMR \cite{Nakai.Y_etal.Phys.-Rev.-B2010}, penetration depth, and thermal conductivity measurements \cite{Hashimoto.K_etal.Phys.-Rev.-B2010}. If the symmetry of the order parameter is $A_{1g}$, line nodes are realized by an accidental crossing of the zero surface of the order parameter and the FSs \cite{Mizukami.Y_etal.Nat.-Commun.2014}, which requires a careful experimental investigation into the location of nodes in the momentum space. 

The structure of the accidental node of the SC gap sensitively depends on the microscopic pairing mechanism. In the spin-fluctuation-mediated pairing mechanism, the SC gap tends to change its sign along the nesting vector. In particular, the pnictogen height sensitively controls the radii of FSs and the nesting vectors, thereby switching the nodal and nodeless SC gaps \cite{Kuroki.K_etal.Phys.-Rev.-B2009}. Suzuki \textit{et al.} \cite{Suzuki.K_etal.Journal-of-the-Physical-Society-of-Japan2011} have calculated the three-dimensional SC gap structure within the random phase approximation (RPA) using the lattice parameters of Ba122P, and  shown that the SC gap on the outer hole FS around the Z [($0, 0, 2\pi$)] point that has a contribution from the $d_{z^{2}}$ orbital character is small (below 2 meV) and may change sign. Furthermore, the gap shows in-plane anisotropy with the strongest tendency to sign change along the direction 45$^{\circ}$ off the Z-X direction. On the other hand, Saito \textit{et al.} \cite{Saito.T_etal.Phys.-Rev.-B2013} have shown that, by taking orbital flucutuations into account, a nodal $s$-wave state may appear with loop-shaped line nodes on the electron FSs. Furthermore, in stark contrast to the spin-fluctuation scenario, the gap function on the outer hole FS around the Z point may have a finite value that is comparable to those in the other two hole FSs, as a result of the interorbital correlations among the $d$ orbitals.

Given the distinct behavior of the theoretical SC gap expected for different pairing interactions, experimental information about the orbital dependence of the SC gap and the location of line nodes will shed light on the microscopic mechanism of the Cooper pairing. In order to clarify them, the SC gap of Ba122P has been studied using various experimental probes. Shimojima \textit{et al.} \cite{Shimojima.T_etal.Science2011,Shimojima.T_etal.Solid-State-Commun.2012} have found a nearly isotropic, orbital-independent SC gap opening on the three hole FSs around the Z point by high-resolution laser-based angle-resolved photoemission spectroscopy (ARPES). Furthermore, Yamashita \textit{et al.} \cite{Yamashita.M_etal.Phys.-Rev.-B2011} observed a fourfold oscillation in angle-resolved thermal conductivity as a function of magnetic field direction within the basal plane. They have concluded that loop-like line nodes appear on the outer electron FS \cite{Khodas.M_etal.Phys.-Rev.-B2012} to explain their observations. However, there is a discrepancy in the literature on the ARPES data utilizing variable energy photons from synchrotron \cite{Zhang.Y_etal.Nat-Phys2012,Yoshida.T_etal.Sci.-Rep.2014}. 


A related system SrFe$_{2}$(As$_{1-x}$P$_{x}$)$_{2}$ (Sr122P) has a slightly higher optimal critical temperature ($x$ = 0.35, $T_{c}$ = 33 K) than that of Ba122P ($x$ = 0.30, $T_{c}$ = 30 K), and its parent compound SrFe$_{2}$As$_{2}$ has a much higher N\'eel temperature ($T_{N}= 197$ K) than that of BaFe$_{2}$As$_{2}$ ($T_{N}$ = 138 K) \cite{Kobayashi.T_etal.Journal-of-the-Physical-Society-of-Japan2014}. While Sr122P and Ba122P share nearly the same pnictogen heights from the square Fe plane, the $c$-axis length is significantly shorter in Sr122P than in Ba122P reflecting the smaller ionic radius of the Sr$^{2+}$ ion than that of the Ba$^{2+}$ ions. In a previous work on as-grown samples of Sr122P \cite{Suzuki.H_etal.Phys.-Rev.-B2014}, we have measured its band dispersions and FSs, and clarified that the outer hole FS is more strongly warped along the $k_{z}$ direction than the corresponding one in Ba122P. The presence of line nodes in Sr122P is also suggested from the $^{31}$P-NMR \cite{Dulguun.T_etal.Phys.-Rev.-B2012}, specific heat \cite{Kobayashi.T_etal.Phys.-Rev.-B2013}, and penetration depth measurements \cite{Murphy.J_etal.Phys.-Rev.-B2013}. It is, therefore, conceivable that the same pairing mechanism gives rise to qualitatively similar nodal SC gaps both in Ba122P and Sr122P and also to some differences arising from the accidental nature of line nodes in the $s$-wave SC state. Thus Sr122P is an ideal system for examining the SC gaps in a systematic way, and new insights into the nodal SC gaps in iron pnictides could be obtained. In particular, the gap function on the more expanded outer hole FS around the Z point may clarify the dominant pairing interaction, because the contribution from the $d_{z^{2}}$ orbital sensitively controls the magnitude of the SC gaps.

\section*{Results}
{\bf Fermi surfaces of SrFe$_{2}$(As$_{0.65}$P$_{0.35}$)$_{2}$.}
In-plane FS mapping taken with $h\nu = 24$ eV and 28 eV is shown in Figs. \arabic{Fig1}(a) and (b), respectively. The first BZ of Sr122P is also shown in the inset of (a). $h\nu =24$ eV corresponds to a plane slightly above the $k_{z}=2\pi$ plane at the zone center (see Fig. \arabic{Fig4}), and $h\nu =28$ eV crosses the X  [($\pi, \pi, 2\pi$)] point, judging from the radius of the hole FSs \cite{Suzuki.H_etal.Phys.-Rev.-B2014} and the elongation of the $\epsilon$ FS along the ($0, 0, 2\pi$)-($\pi, \pi, 2\pi$) line \cite{Yamashita.M_etal.Phys.-Rev.-B2011}. There are three hole-like bands crossing $E_{F}$ around the Z point, which form three circular FSs, and two electron-like bands crossing $E_{F}$ around the X point. These FS topologies are the same as those for the as-grown samples \cite{Suzuki.H_etal.Phys.-Rev.-B2014}, demonstrating that the annealing process (described in Ref. \onlinecite{Kobayashi.T_etal.Journal-of-the-Physical-Society-of-Japan2014}) does not modify the electronic structure significantly. 
The dominant orbital character of the hole bands crossing $E_{F}$ are $d_{xz/yz}$ ($\beta$,$\gamma$) around the $\Gamma$ point, and  $d_{xy}$ ($\alpha$), $d_{xz/yz}$ ($\beta$), and $d_{xz/yz}$ and $d_{z^{2}}$ ($\gamma$) around the Z point. The orbital character for electron bands is $d_{xz/yz}$ ($\delta$) and $d_{xy}$ ($\epsilon$).

{\bf Superconducting gaps in the hole Fermi surfaces.}
Now we investigate the SC gap in the momentum space. Figure \arabic{Fig2}(a) shows FS mapping taken with the 7 eV laser. The observation of three FSs and their radii indicate that $h\nu=7$ eV measures a plane close to the Z point. We define the FS angle $\theta$ with respect to the nearest-neighbor Fe-Fe direction as indicated in the figure. Symmetrized energy distribution curves (EDCs) below and above $T_{c}$ for the three FSs are shown in Figs. \arabic{Fig2}(b)-(d). The closure of the gap above $T_{c}$ demonstrates that the gap originates from superconductivity. To extract the gap size from the data, symmetrized EDCs below $T_{c}$ have been fitted to a
phenomenological low-energy spectral function of the form \cite{Norman.M_etal.Phys.-Rev.-B1998}
\begin{equation}
A(k_{F},\omega)=-\frac{1}{\pi}\frac{\Sigma^{\prime\prime}(k_{F},\omega)}{[\omega-\Sigma^{\prime}(k_{F},\omega)]^{2}+[\Sigma^{\prime\prime}(k_{F},\omega)]^{2}},
\end{equation}
with the self-energy at $k_{F}$,
\begin{equation}
\Sigma(k_{F},\omega)=-i\Gamma_{1}+\Delta^{2}/[\omega+i\Gamma_{0}].
\end{equation}
Here, $\Gamma_{1}$ is the single-particle scattering rate, $\Delta$ is the SC gap, and $\Gamma_{0}$ is the inverse pair lifetime. It is known that a finite $\Gamma_{0}$ well reproduces the spectral broadening in the pseudogap state of the cuprates \cite{Norman.M_etal.Phys.-Rev.-B1998}. Although the importance of the $\Gamma_{0}$ term has been pointed out \cite{Khodas.M_etal.Phys.-Rev.-B2012}, here we have set $\Gamma_{0}=0$ because the present symmetrized EDCs do not have maxima at $\omega=0$ and the EDCs at 40 K (just above $T_{c}=33$ K) in panels (b)-(d) do not show any signature of pseudogap. When the spectra have clear coherence peaks and deep gaps at $\omega=0$ as in the $\alpha$ and $\beta$ FSs, this fitting procedure works well and the gap size thus deduced yields the coherence peak position [panels (b) and (c)] . However, in the absence of clear SC peaks due to the small gap or low quasiparticle weight as in the case of the $\gamma$ FS [panel (d)], the fitting is quite unstable and a wide range of parameter sets can reproduce the spectral lineshapes, making it difficult to estimate the SC gap values. Therefore, we have estimated the SC gap of the $\gamma$ FS from the shoulder structures of the low-$T$ spectra if they exist, as indicated by open circles in Fig. \arabic{Fig2}(d), and we plot only error bars if there is no discernible shoulder structure. We had to resort to visual inspection of the spectra to determine the shoulder positions with larger error bars, but this is partly justified by the fact that the shoulder positions in Fig. \arabic{Fig2}(d) are in good agreement with the energy at which the changes between the 3 K and 40 K spectra occurs, evidencing that the shoulders are indeed caused by the SC gap opening. We notice that the shoulder positions exhibit an angular dependence, and that the shoulders are almost absent in the -2$^{\circ}$ and -15$^{\circ}$ spectra. Note that even when there is no signature of SC gap opening, a positive $\Delta$ with comparable $\Gamma_{1}$ can reproduce spectra like the -15$^{\circ}$ one in panel (d), as pointed out by Khodas \textit{et al.} \cite{Khodas.M_etal.Phys.-Rev.-B2012}. Furthermore, the spectral intensity of the $\gamma$ band is weaker than that in the other bands. Thus, throughout this paper, we do not claim the presence of nodes simply based on the absence of a gap, as previously done for Ba122P by Zhang \textit{et al.} \cite{Zhang.Y_etal.Nat-Phys2012}. Instead, our argument is based on the anisotropy of the shoulder structures. 
The extracted SC gap is plotted in Fig. \arabic{Fig2}(e) as a function of $\theta$ and the dotted lines indicate the fitting of the $\alpha/\beta$ FS data by $\Delta(\theta)=\Delta_{0}+\Delta_{2}\cos(4\theta)$. The $\alpha$ and $\beta$ FSs have isotropic gaps of $\Delta_{\alpha}(\theta)=5.1+0.4\cos(4\theta)$ meV and $\Delta_{\beta}(\theta)=4.5+0.3\cos(4\theta)$ meV, respectively. On the other hand, the $\gamma$ FS shows a signature of gap anisotropy with the minimum around $\theta=0^{\circ}$ and the maxima around $\theta=45^{\circ}$.

In order to study a wider momentum range, we employed synchrotron ARPES. Figure \arabic{Fig3}(a) shows symmetrized energy-momentum ($E$-$k$) ARPES intensity plots for the hole FSs taken with $h\nu =23$ eV photons, which again corresponds to the $k_{z}=2\pi$ plane including the Z point at the zone center. The $k_{F}$ positions are indicated by arrows, and the momentum cuts in the BZ are illustrated in Fig. \arabic{Fig3}(b). As we have seen in the laser data, the SC coherence peak intensity is the strongest for the $\alpha$ sheet, and the weakest for the $\gamma$ sheet. The symmetrized EDCs are shown in Figs. \arabic{Fig3}(c)-(e). We observe more pronounced SC coherence peaks for the $\alpha$ FS [panel (c)] than the laser data, probably reflecting larger photoemission cross sections at $h\nu=23$ eV. For the $\beta$ FS [panel (d)], the spectral shape is similar to the laser data. For the $\gamma$ FS [panels (e1) and (e2)], we again observe an indication of anisotropy from the shoulders indicated by the circles.  The estimated SC gap is plotted in Fig. \arabic{Fig3}(f). We again obtain the isotropic gaps of $\Delta_{\alpha}(\theta)=7.5+0.1\cos(4\theta)$ meV for the $\alpha$ FS and $\Delta_{\beta}(\theta)=6.1+0.3\cos(4\theta)$ meV for the $\beta$ FS. For the $\gamma$ FS, the amplitude of the anisotropic shoulder is $\sim$ 5 meV around $\theta=\pm 45^{\circ}$ and $135^{\circ}$, and a signature of gap minima around $\theta=90^{\circ}$ (equivalent to $\theta=0^{\circ}$), consistent with the laser data. However, we also observe flat symmetrized EDCs between $-14^{\circ}<\theta<49^{\circ}$ in Figs. \arabic{Fig3}(e1) and (e2), probably because the momentum cuts approach the FS edge, which tends to reduce the photoemission intensity and leads to featureless EDCs.

In order to study the $k_{z}$ dependence of the SC gap, we have performed measurements by varying photon energy. Figure  \arabic{Fig4}(a) shows $E$-$k$ intensity plots for the hole FSs taken using various photon energies $h\nu=23$-31 eV. The momentum cuts are parallel to those in Fig. \arabic{Fig3}. At the zone center, $h\nu=$23 eV corresponds to the $Z$ point and $h\nu=$31 eV to the $\Gamma$ point. We observe a strong warping of the $\gamma$ FS as $h\nu$ is varied, as illustrated in Fig. \arabic{Fig4}(b). Symmetrized EDCs are plotted in Figs. \arabic{Fig4}(c)-(e) and the estimated gap magnitudes are plotted in Fig. \arabic{Fig4}(f). As $k_{z}$ moves from the Z point toward the $\Gamma$ point, the gap on the $\alpha$ FS becomes smaller and the coherence peak intensity also becomes lower, the line shape for the $\beta$ FS is almost unchanged, and the gap of the $\gamma$ FS slightly increases.

{\bf Superconducting gaps in the electron Fermi surfaces.}
Next we move to the electron FSs. Figure \arabic{Fig5}(a) shows symmetrized $E$-$k$ intensity plots for the electron FSs taken with $h\nu=24$ eV photons. It can be seen that the SC gap is finite for all the $k_{F}$'s. Judging from the fact that the two electron FSs are almost circular around the $(\pi, \pi)$ point as illustrated in Fig. \arabic{Fig5}(b), $h\nu=24$ eV would correspond to a plane with $k_{z}\sim\pi$ . The symmetrized EDCs are shown in Figs. \arabic{Fig5}(c) and (d). We observe SC coherence peaks in the $\delta$ sheet. For the $\epsilon$ sheet, although the background level is high due to the presence of the $\delta$ band below the $\epsilon$ band, we still observe a gap at $\omega=0$. The estimated SC gap is plotted in Fig. \arabic{Fig5}(e). Both the $\delta$ and $\epsilon$ FSs have the isotropic SC gaps of $\Delta_{\delta}(\theta)=7.0+0.2\cos(2\theta)+0.6\cos(4\theta)$ meV and $\Delta_{\epsilon}(\theta)=7.8+0.0\cos(2\theta)+0.4\cos(4\theta)$ meV. This isotropic gap is in contrast to the observed line nodes \cite{Yamashita.M_etal.Phys.-Rev.-B2011} and gap anisotropy \cite{Yoshida.T_etal.Sci.-Rep.2014} in the electron pockets of Ba122P.

To investigate the $k_{z}$ dependence of the gap on the electron FSs, we also performed in-plane measurements with another $h\nu$ of 28 eV. Figure \arabic{Fig6}(a) shows symmetrized $E$-$k$ intensity plots. $h\nu=28$ eV corresponds to the $k_{z}=2\pi$ plane including the X point, as illustrated in Fig. \arabic{Fig6}(b). One again observes that the SC gap is finite for all the $k_{F}$'s. The symmetrized EDCs are shown in Figs. \arabic{Fig6}(c) and (d). The line shapes are close to those in Fig. \arabic{Fig5}. The SC gap is plotted in Fig. \arabic{Fig6}(e). Again one obtains almost isotropic gaps of $\Delta_{\delta}(\theta)=7.5+0.4\cos(2\theta)+0.4\cos(4\theta)$ meV and $\Delta_{\epsilon}(\theta)=8.3+0.7\cos(2\theta)+0.0\cos(4\theta)$ meV. From these results, one may exclude the possibility of nodes on the electron FSs.


\section*{Discussion}
Although it was not possible to determine the three-dimensional nodal structure on the entire FSs from the limited momentum cuts in our measurements, line nodes are most likely located near crossing point of the $\theta=0^{\circ}$ line on the $\gamma$ hole FS around the Z point considering the anisotropic small gap and very weak coherence peaks. As for the shape of the node, vertical line nodes in the entire $k_{z}$ seem unlikely, considering the increase of the gap on the $\gamma$ FS near the $\Gamma$ point, as shown in Fig. \arabic{Fig4}(e). Based on these considerations, if we assume that the order parameter has the $A_{1g}$ symmetry,  the most probable scenario is loop-like nodes on the $\gamma$ hole FS, around the Z point, and around the $\theta=0^{\circ}$ line (the Z-X line), as illustrated in Fig. \arabic{Fig7}(b) (details below). We now examine whether an order parameter $\Delta ({\bm k})$ with $A_{1g}$ symmetry smoothly varying in ${\bm k}$ space can explain our three-dimensional experimental data and produce loop nodes. For the sake of minimizing the number of free parameters, we assume that the order parameter $\Delta (k_{x},k_{y})$ is $k_{z}$-independent and that the $k_{z}$ dependence of the SC gap is solely due to the warping of the $\gamma$ FS along $k_{z}$. The observed anisotropy of the SC gaps can thus be reasonably reproduced by including four Fourier components: $\Delta ({\bm k})=1+4[\cos(k_{x}a)+\cos(k_{y}a)]-7\cos(k_{x}a)\cos(k_{y}a)+4[\cos(2k_{x}a)+\cos(2k_{y}a)]$ (meV) [Fig. \arabic{Fig7}(a)]. On the $\gamma$ FS, this $\Delta ({\bm k})$ changes sign around the Z-X line as illustrated in Fig. \arabic{Fig7}(b), and accordingly produces small loop-like nodes. Figures \arabic{Fig7}(c) and (d) show comparison between this $\Delta ({\bm k})$ and the in-plane experimental data for the $\gamma$ FS [Fig. \arabic{Fig2}(e) and Fig. \arabic{Fig3}(f)].  $\Delta ({\bm k})$ well reproduces the angular dependence of the data. Possible experimental $k_{z}$ broadening would further improve the fit. As shown in Fig. \arabic{Fig7}(e), the gradual increase of the gap along the $k_{z}$ direction [Fig. \arabic{Fig4}(f)] is also well reproduced.

Here we discuss the comparison of the present result with the theoretical calculations in the literature. Suzuki \textit{et al.} \cite{Suzuki.K_etal.Journal-of-the-Physical-Society-of-Japan2011} calculated the gap structure of Ba122P based on the spin-fluctuation exchange mechanism and showed that sign reversal occurs on the $\gamma$ FS around the Z point, consistent with the present result. Quantitatively, however, their result shows that the sign-reversed gap on the $\gamma$ FS has a (negative) maximum on the Z-X line, which is by 45$^{\circ}$ rotated from the present result. On the other hand, Saito \textit{et al.} \cite{Saito.T_etal.Phys.-Rev.-B2013} showed that the SC gap on all the hole FS becomes of the same sign, the gap on the electron FSs becomes anisotropic, and nodes may be formed on the electron FSs depending on the magnitude of microscopic parameters. Also, the gap of the hole FSs on the $k_{z}=\pi$ plane is the largest for the $\gamma$ FS. Since the $\gamma$ FS in Sr122P is more strongly warped than Ba122P due to the smaller $c$-axis lattice parameter, the theoretical SC gap structure on the $\gamma$ FS of Sr122P could be largely modified from those of Ba122P. A direct calculation of Sr122P using its lattice parameters would further facilitate the understanding of microscopic pairing mechanism both in Ba122P and Sr122P.


Finally, we would like to make a remark on the effect of impurities on the SC gap structure. A recent study on the effects of electron irradiation and natural disorder in Sr122P single crystals \cite{Strehlow.C_etal.Phys.-Rev.-B2014} showed that the low-temperature behavior of the penetration depth can be described by a power-law function $\Delta \lambda (T)=AT^{n}$, with $n$ close to one for pristine annealed samples and larger than two for electron-irradiated ones, implying that the nodes are accidental and can be lifted by the introduction of disorder. For Ba122P, Mizukami \textit{et al.} \cite{Mizukami.Y_etal.Nat.-Commun.2014} also found that the nodal state changes to a nodeless state showing fully gapped excitations by introducing nonmagnetic point defects by electron irradiation. Theoretically, Kontani \textit{et al.} \cite{Kontani.H_etal.Phys.-Rev.-Lett.2010} proposed a possibility of crossover from $s^{\pm}$ wave to $s^{++}$ wave with increasing the impurity concentrations. The present annealed samples, which yielded $n\sim 1$ in Strehlow \textit{et al.}'s penetration depth measurement \cite{Strehlow.C_etal.Phys.-Rev.-B2014}, can be reasonably regarded as ``pure'', and a comparison with theory with low impurity concentration is justified. To further clarify this issue, direct ARPES investigation into the change of SC gap structure with different impurity concentrations is left for future studies.

In conclusion, we have measured the SC gaps of optimally-doped Sr122P in wide three-dimensional momentum space by laser-based and synchrotron-radiation ARPES. Around the Z point, the $\alpha$ and $\beta$ hole FSs have isotropic SC gap of $8$ meV and $6$ meV, respectively. The gap on the $\gamma$ FS is anisotropic with the amplitude of $5$ meV with minima located along the Z-X direction. The electron FSs have isotropic gap of $8$ meV, which is almost independent of the $k_{z}$. As for the location of line nodes, we propose that there exists a sign change on the $\gamma$ hole FS, around the Z point, and around the Z-X line. A direct theoretical calculation of the Sr122P SC gap structure using its lattice parameters, together with the existing data on Ba122P and Sr122P, would further deepen our understanding of the nodal superconductivity in the iron pnictides.

\section*{Acknowledgements}
We are grateful to M. Nakajima and H. Eisaki for enlightening discussions.
Stanford Synchrotron Radiation Lightsource is operated by the Office of Basic Energy Science,
US Department of Energy.  H.S. and M.Horio acknowledge financial support from Advanced Leading Graduate Course for Photon Science (ALPS) and the JSPS Research Fellowship for Young Scientists.

\section*{Author contributions}
H.S. and K.O. performed Laser ARPES experiment with the help of Y.O., H.Y. and S.S.. H.S., K.O., T.Y., M.Horio and L.C.C.A. performed synchrotron ARPES experiment with the help of M.Hashimoto., D.H.L. and Z.X.S.. T.K., S.M. and S.T. grew the Sr122P single crystals and performed sample characterizations. H.S. analyzed the experimental data. H.S. and A.F. wrote the manuscript with comments from all co-authors.

\section*{Competing interests}
The authors declare no competing interests.

\section*{Data availability}
The data sets generated during and/or analysed during the current study are available from the corresponding author on reasonable request.

\newpage
\section*{Methods}
Sr122P single crystals were prepared by the self-flux method described in Ref. \onlinecite{Kobayashi.T_etal.Phys.-Rev.-B2013} and postannealed to achieve the optimal $T_{c}$ of 33 K \cite{Kobayashi.T_etal.Journal-of-the-Physical-Society-of-Japan2014}. The $T_{c}$ was determined by the onset of Meissner diamagnetic signal with the transition width of $\Delta T_{c}\simeq$ 5 K.  ARPES experiments were carried out at beamline 5-4 of Stanford Synchrotron Radiation Lightsource (SSRL) and at a laser ARPES apparatus at Institute for Solid State Physics (ISSP). In order to obtain clean surfaces, samples were cleaved \textit{in situ} at pressure better than $1\times10^{-10}$ Torr. Cleavage occurs along the $ab$ planes. The kinetic energies and momenta of photoelectrons were measured using Scienta R4000 electron energy analyzers. In the following, the $x$ and $y$ axes point from Fe towards the nearest neighbor Fe atoms, the $X$ and $Y$ axes point from the Fe atom towards the second nearest-neighbor Fe atoms, and the $z$ axis is parallel to the $c$-axis. In-plane ($k_{X}$, $k_{Y}$) and out-of-plane momenta ($k_{z}$) are expressed in units of $1/a$ and $1/c$, respectively, where $a=3.90$ \AA \,and $c=12.09$ \AA\, are the in-plane and out-of-plane lattice constants. Calibration of the Fermi level ($E_{F}$) was achieved using the spectra of gold which was in electrical contact with the samples. Incident photons from 21 eV to 35 eV from the synchrotron were linearly $p$-polarized. For laser ARPES, the incident photons with $h\nu=6.998$ eV were linearly $s$-polarized. The total energy resolution was $\Delta E\sim$ 6 meV at $T=5$ K for synchrotron-radiation ARPES, and $\Delta E\sim$ 1 meV at $T=3$ K for laser ARPES.

\newpage

\newpage

\begin{figure}[htbp]
   \centering 
   \includegraphics[width=14cm]{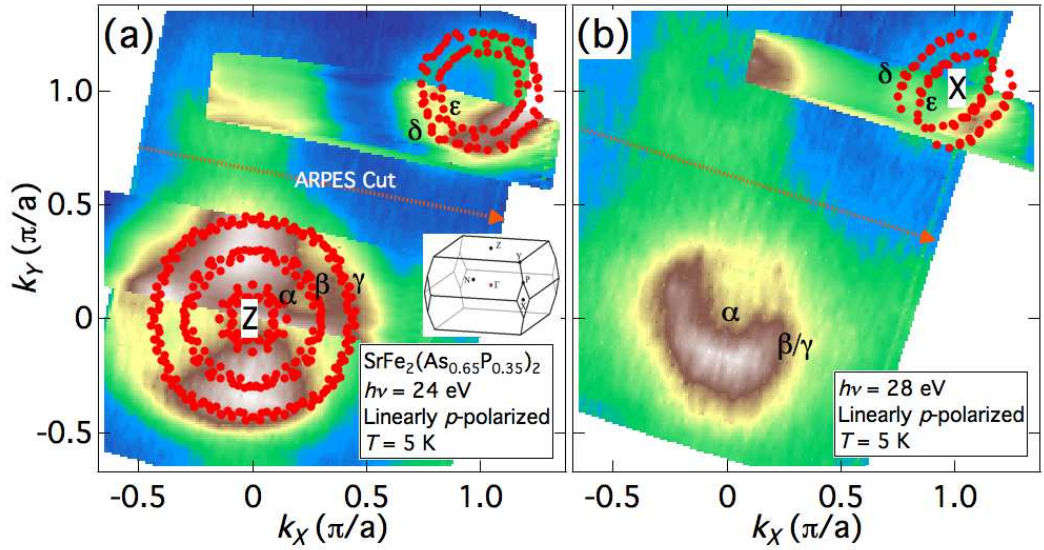}
   \caption*{\raggedright\textbf{Figure 1 $\vert$ Fermi surfaces of SrFe$_{2}$(As$_{0.65}$P$_{0.35}$)$_{2}$.} (a),(b) In-plane Fermi surface (FS) mapping for SrFe$_{2}$(As$_{1-x}$P$_{x}$)$_{2}$ (Sr122P) ($x=0.35$) taken with $h\nu=24$ eV [(a)] and 28 eV [(b)]. Filled circles indicate the Fermi wave vectors ($k_{F}$'s) and their symmetrized points. The first Brillouin zone of Sr122P is shown in the inset of panel (a).}
  \label{GapFS}
 \end{figure}

\newpage

\begin{figure}[htbp]
   \centering 
   \includegraphics[width=14cm]{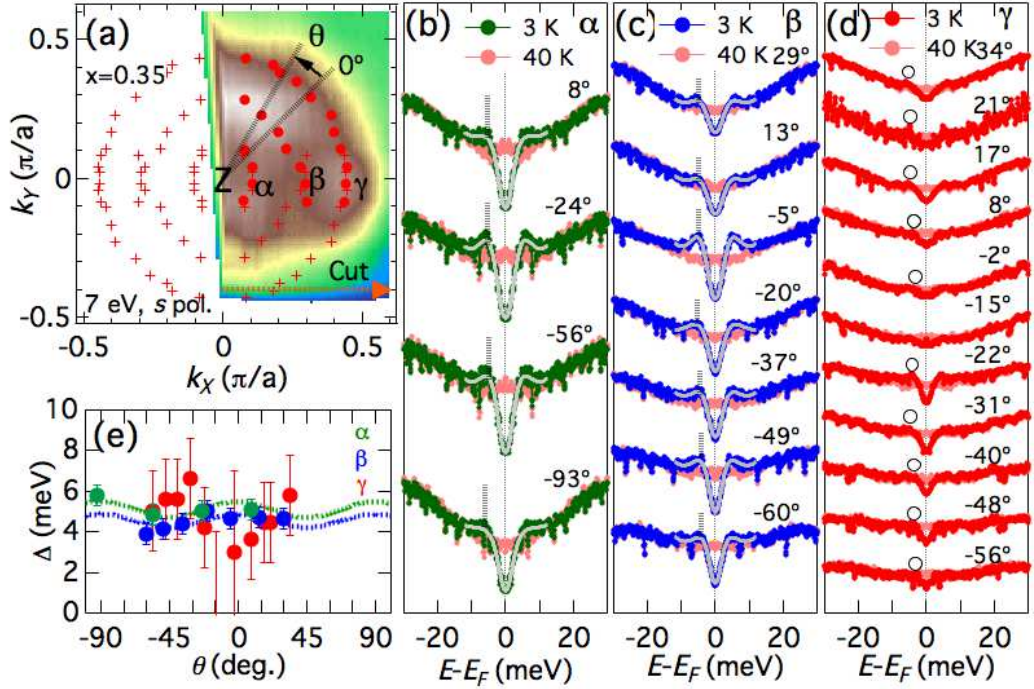}
   \caption*{\raggedright\textbf{Figure 2 $\vert$ Superconducting gaps in the hole FSs investigated with Laser ARPES.} (a) FS mapping of the hole FSs taken with $h\nu=7$ eV laser. Red circles indicate the $k_{F}$ positions, where the superconducting (SC) gap sizes are estimated, and crosses show their symmetrized points. FS angle $\theta$ is defined with respect to the Fe-Fe direction as indicated in the figure. (b)-(d) Symmetrized energy distribution curves (EDCs) at $k_{F}$'s for the inner ($\alpha$), middle ($\beta$), and outer ($\gamma$) FSs, respectively. Blue and red spectra show data taken at 3 K and 40 K, respectively. Gray curves indicate fitted curves. Vertical bars indicate estimated SC gaps. For the $\gamma$ FS, where the SC coherence peaks are almost absent, we determine SC gap from the shoulder structures of the spectra, as indicated by circles. (e) SC gap $\Delta$ plotted as a function of FS angle $\theta$. The dotted lines indicate fitting by $\Delta(\theta)=\Delta_{0}+\Delta_{2}\cos(4\theta)$.}
  \label{lasergap}
 \end{figure}

\newpage

\begin{figure}[htbp]
   \centering 
   \includegraphics[width=14cm]{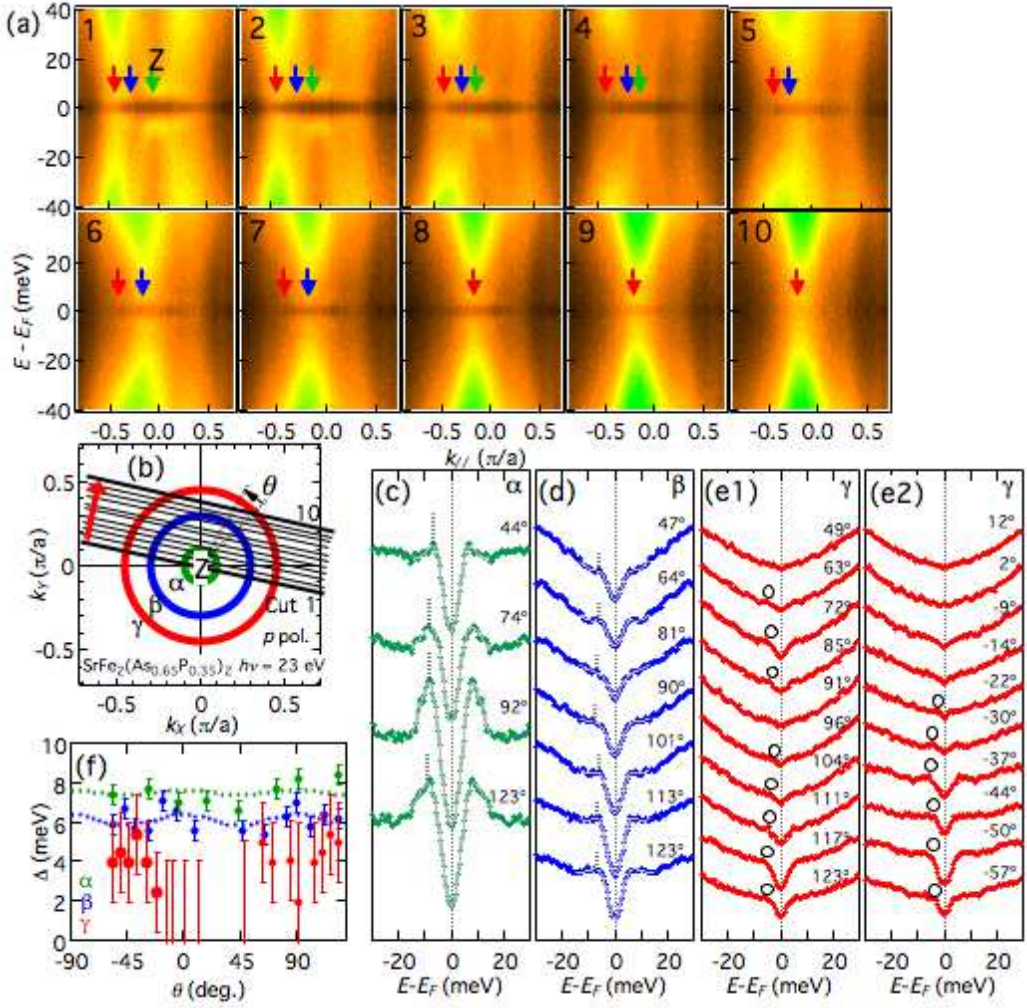}
   \caption*{\raggedright\textbf{Figure 3 $\vert$ Superconducting gaps in the hole FSs around the Z point.} (a) Symmetrized energy-momentum ($E$-$k$) intensity plots at 5 K for the hole FSs taken with $h\nu=23$ eV photons. The $k_{z}$ value corresponds to that of the Z point at the zone center. The location of each cut is indicated in panel (b). The $k_{F}$ positions are shown by arrows. (b) Schematic figure of the hole FSs. (c)-(e) Symmetrized EDCs. Estimated gap is indicated by vertical lines [(c), (d)] and circles [(e1), (e2)]. (f) SC gap as a function of FS angle $\theta$. The dotted lines indicate fitting by $\Delta(\theta)=\Delta_{0}+\Delta_{2}\cos(4\theta)$.}
  \label{IntHole}
 \end{figure}
 
\newpage

\begin{figure}[htbp]
   \centering 
   \includegraphics[width=14cm]{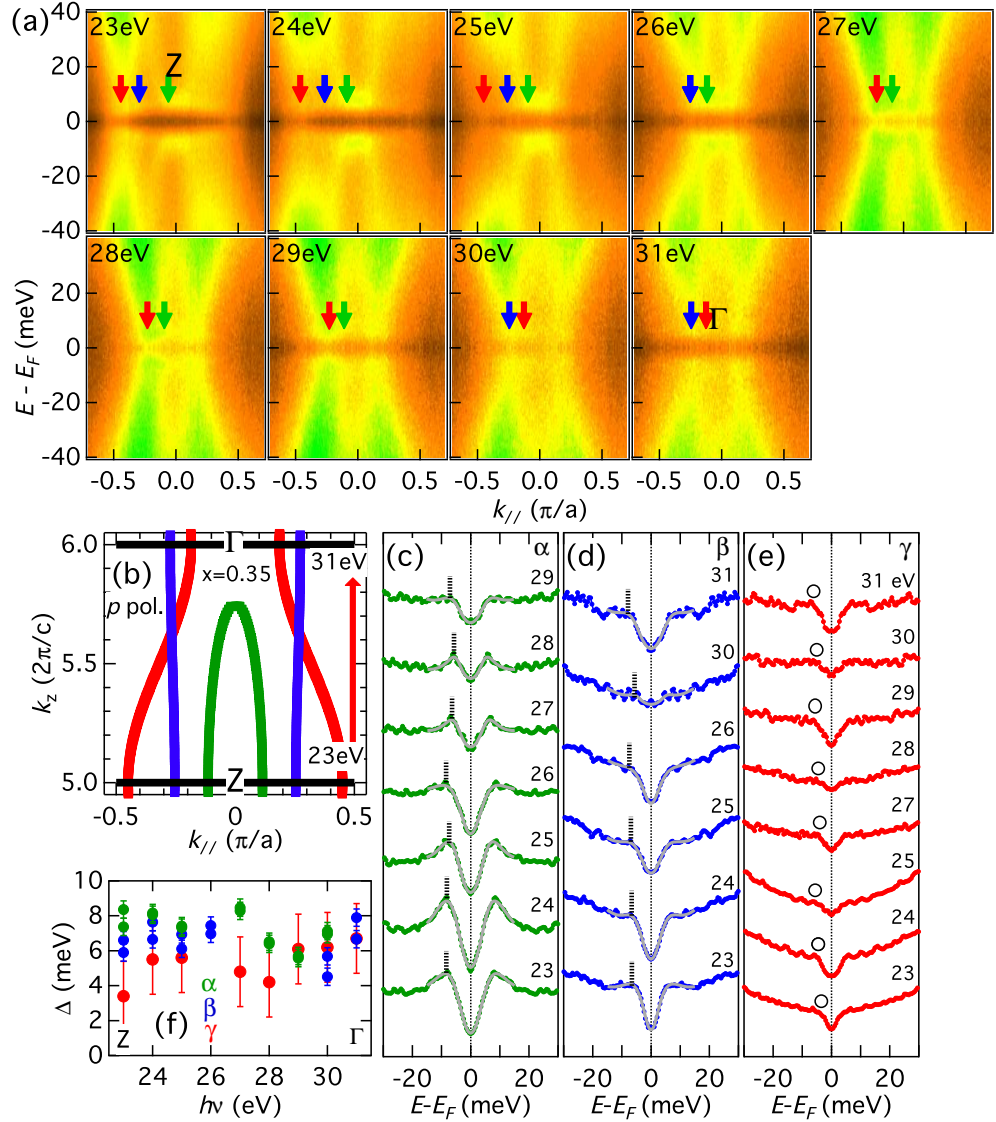}
   \caption*{\raggedright\textbf{Figure 4 $\vert$ Superconducting gaps in the hole FSs along the $k_{z}$ direction.} (a) Symmetrized $E$-$k$ intensity plots for the hole FSs taken along the $k_{z}$ direction using $h\nu=23$-31 eV photons. (b) Schematic cross-sections of hole FSs in the $k_{\parallel}$-$k_{z}$ plane and momentum cuts, which are parallel to those in Fig. \arabic{Fig3}(b). (c)-(e) Symmetrized EDCs. Estimated gap is indicated by vertical lines [(c), (d)] and circles [(e)]. (f) SC gap magnitudes as functions of incident photon energy.}
  \label{IntHolekz}
 \end{figure}

\newpage

\begin{figure}[htbp]
   \centering 
   \includegraphics[width=14cm]{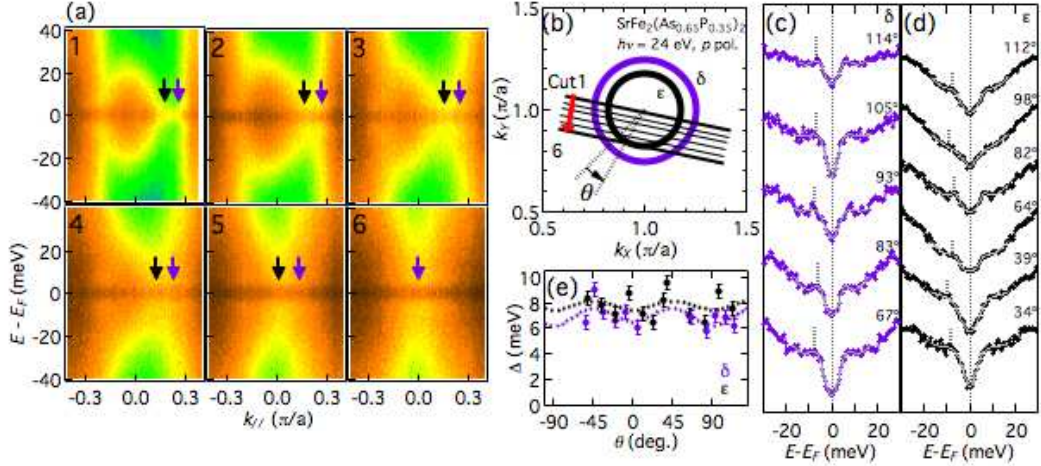}
   \caption*{\raggedright\textbf{Figure 5 $\vert$ Superconducting gaps in the electron FSs probed with 24eV photons.} (a) Symmetrized $E$-$k$ intensity plots for the electron FSs taken with $h\nu=24$ eV photons. (b) Schematic figure of the electron FSs. (c), (d) Symmetrized EDCs.  Estimated gap is indicated by vertical lines.
 (e) SC gap as a function of FS angle $\theta$. The dotted lines indicate fitting by $\Delta(\theta)=\Delta_{0}+\Delta_{1}\cos(2\theta)+\Delta_{2}\cos(4\theta)$.}
  \label{Intele}
 \end{figure}

\newpage

\begin{figure}[htbp]
   \centering 
   \includegraphics[width=14cm]{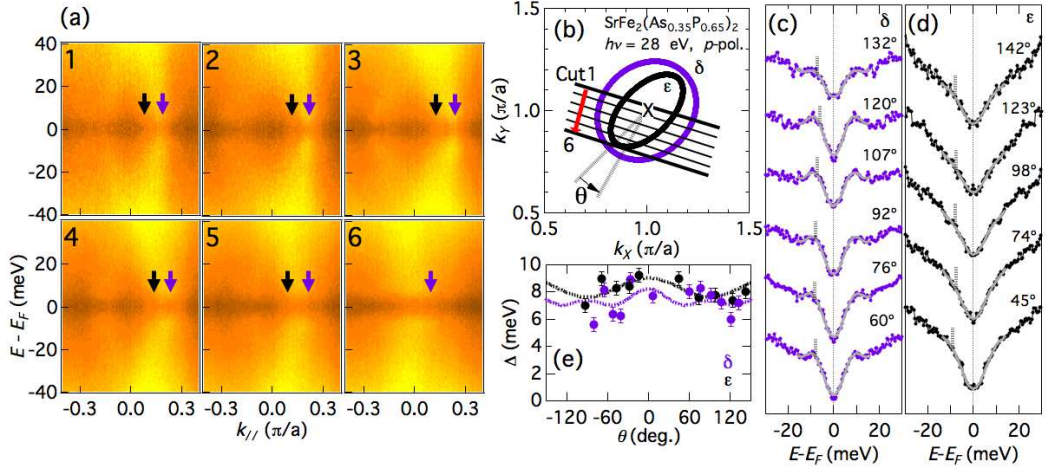}
   \caption*{\raggedright\textbf{Figure 6 $\vert$ Superconducting gaps in the electron FSs around the X point.} (a) Symmetrized $E$-$k$ intensity plots for the electron FSs taken with $h\nu=28$ eV photons. (b) Schematic figure of the electron FSs. The $k_{z}$ at ($\pi,\pi$) for 28 eV corresponds to that of the X point. (c), (d) Symmetrized EDCs. (e) SC gap as a function of $\theta$. The dotted lines indicate fitting by $\Delta(\theta)=\Delta_{0}+\Delta_{1}\cos(2\theta)+\Delta_{2}\cos(4\theta)$.}
  \label{Intele28}
 \end{figure}

\newpage

\begin{figure}[htbp]
   \centering 
   \includegraphics[width=14cm]{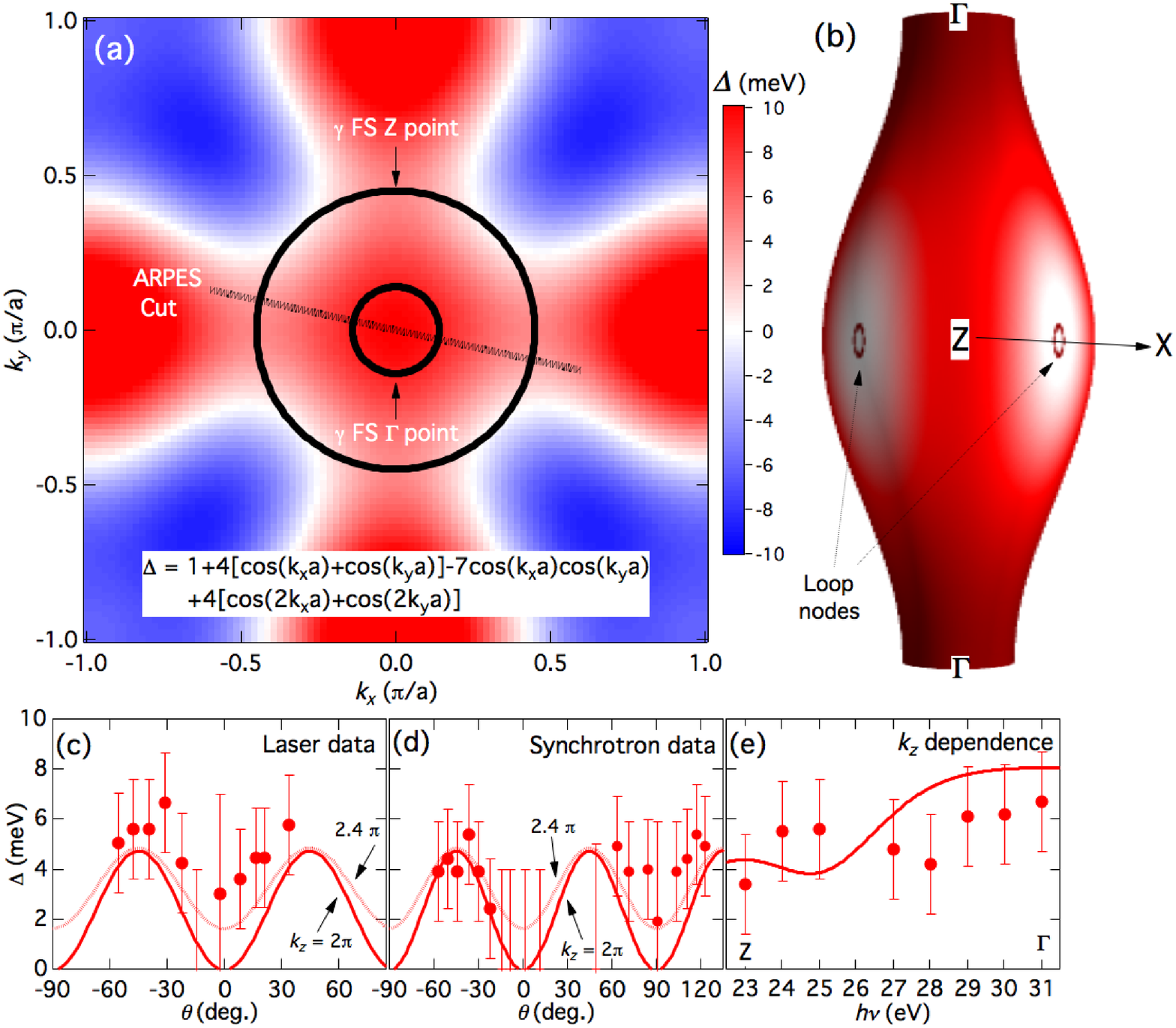}
   \caption*{\raggedright\textbf{Figure 7 $\vert$ Fitting of the nodal superconducting gap structure of the $\gamma$ hole FS.} (a) A color plot of $k_{z}$-independent SC order parameter $\Delta ({\bm k})=1+4[\cos(k_{x}a)+\cos(k_{y}a)]-7\cos(k_{x}a)\cos(k_{y}a)+4[\cos(2k_{x}a)+\cos(2k_{y}a)]$ (meV) for the $\gamma$ FS. The FS cross sections around the $\Gamma$ and Z points are shown by circles and the ARPES momentum cut is indicated by a dotted line. (b) Three-dimensional illustration of $\Delta ({\bm k})$ on the $\gamma$ FS. Line nodes are formed around the Z-X direction. (c)-(e) Comparison of $\Delta ({\bm k})$ with the experimental data. In panels (c) and (d), in order to take into account possible $k_{z}$ broadening in the experimental data, the values of $\Delta ({\bm k})$ both on $k_{z}=2\pi$ and $2.4\pi$ planes are plotted.}
  \label{gapfit}
 \end{figure}

\end{document}